\documentclass[twocolumn,aps,floatfix,superscriptaddress]{revtex4}
\pdfoutput=1 
\usepackage{amsmath,amssymb,eucal,graphicx,float,epstopdf}
\begin{document}
\title{Kinetics of Aggregation with Choice}
\author{E.~Ben-Naim}
\affiliation{Theoretical Division and Center for Nonlinear Studies,
Los Alamos National Laboratory, Los Alamos, New Mexico 87545, USA}
\author{P.~L.~Krapivsky}
\affiliation{Department of Physics, Boston University, Boston, Massachusetts 02215, USA}
\begin{abstract}
We generalize the ordinary aggregation process to allow for choice. In
ordinary aggregation, two random clusters merge and form a larger
aggregate.  In our implementation of choice, a target cluster and two
candidate clusters are randomly selected, and the target cluster
merges with the larger of the two candidate clusters. We study the
long-time asymptotic behavior, and find that as in ordinary
aggregation, the size density adheres to the standard scaling form.
However, aggregation with choice exhibits a number of novel
features. First, the density of the smallest clusters exhibits
anomalous scaling.  Second, both the small-size and the large-size
tails of the density are overpopulated, at the expense of the density
moderate-size clusters. We also study the complementary case where the
smaller candidate clusters participates in the aggregation process,
and find abundance of moderate clusters at the expense of small and
large clusters.  Additionally, we investigate aggregation processes
with choice among multiple candidate clusters, and a symmetric
implementation where the choice is between two pairs of clusters.
\end{abstract}
\maketitle

\section{Introduction}
The concept of choice plays a central role in queuing theory,
algorithms, and computer science \cite{broder,adler,michael}.  In
particular, the so-called ``power of choice'' has been widely explored
in the Achlioptas processes that models evolution of random graphs
\cite{ASS}.  An intriguing, apparently discontinuous, percolation
transition, termed ``explosive percolation'', has been observed in
numerical studies of the original Achlioptas process and several of
its variants \cite{ASS,Ziff09,FL09,KKK09,RF09,DM10,RF10,van10}.
However, it was later shown that this transition, albeit unusually
steep, is actually continuous \cite{Dor10,Maya10,RW11,J11}.

The presence of choice can lead to lack of self-averaging
\cite{RW12,Raissa14}, truly discontinuous percolation transitions, and
multiple giant components \cite{Frieze04,Raissa11}. The power of
choice has been also studied in the realm of growing networks
\cite{Raissa07,KR14}, and it has been shown that it leads to phase
transitions, including the emergence of a macroscopic hub
\cite{KR14}. The classical evolving random graph model \cite{ER} is
equivalent to an aggregation process in which clusters merge with rate
equal to the product of their masses \cite{Ald99,FL03,book}. Yet, 
theoretical analysis of this aggregation process with choice has
proven largely elusive \cite{ASS,Dor10,RW11}.

In this study, we generalize the most basic aggregation process
\cite{Ald99,FL03,book} to include choice. While a complete theoretical
description in the form of the explicit cluster-size density appears
to be out of reach, many features of this distribution can be
understood analytically. In particular, we find the density of the
smallest clusters and the tails of the size distribution. In general,
we demonstrate how choice can be used to control the size
distribution.

In ordinary aggregation, two clusters are chosen at random and are
joined to form a larger cluster. To incorporate choice, we alter this
aggregation process by randomly selecting one target cluster and two
candidate clusters. The target cluster merges with the larger of the
two candidate clusters, leaving the smaller of the two candidate
clusters unaffected. Starting with a uniform size distribution, this
elementary aggregation event is repeated indefinitely.

\begin{figure}
\centering
\includegraphics[width=5cm]{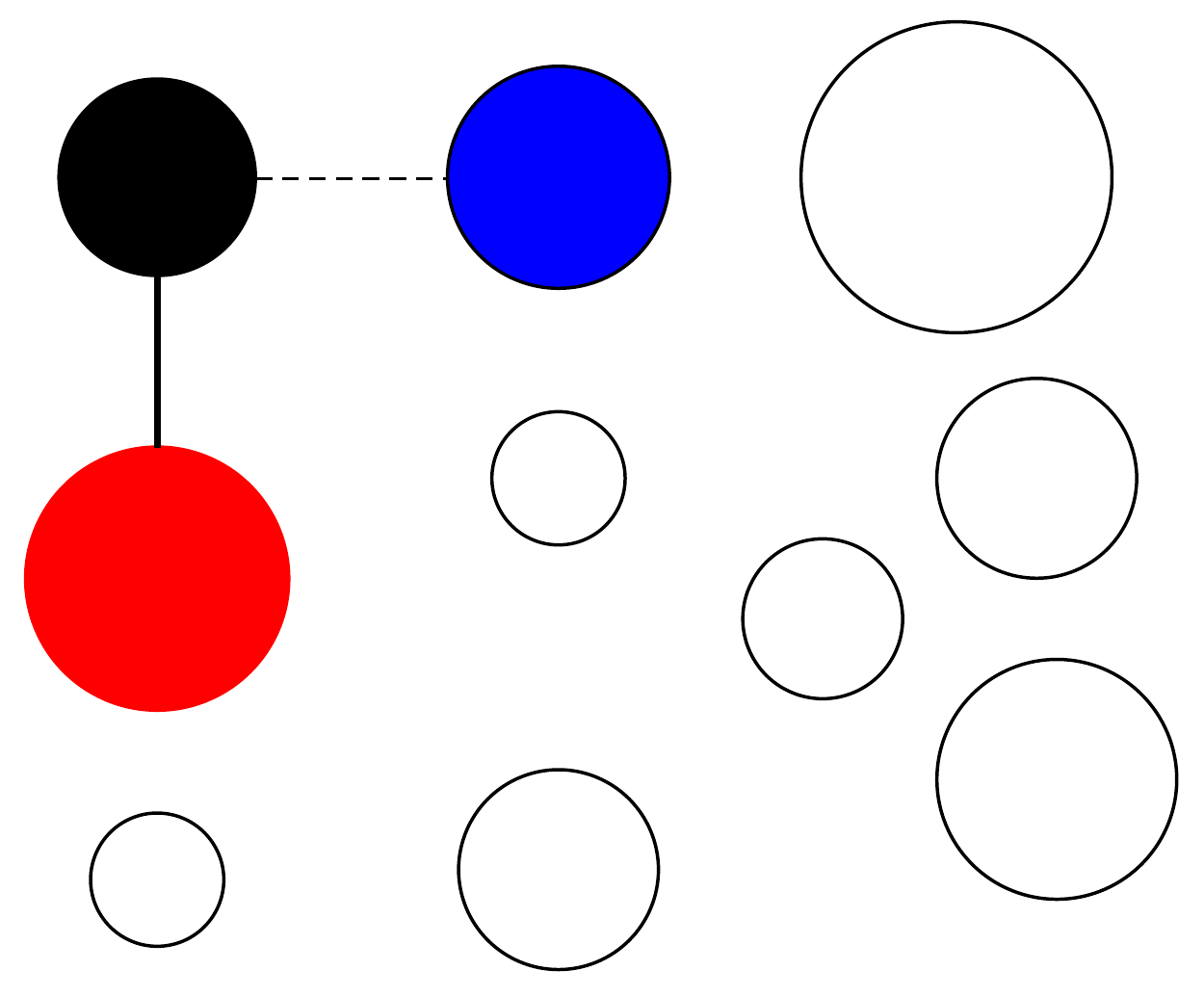}
\caption{Illustration of the aggregation process with choice. Clusters
  are shown as disks, the bigger the disk the larger its size. The
  target cluster (black disk) and two potential merging partners 
  (red and blue disks) are randomly drawn. The larger cluster (red
  disk) is chosen as the actual merging partner in the maximal choice
  case. In the minimal choice case, the smaller cluster (blue disk) is
  chosen.}
\label{fig-max}
\end{figure}

We study kinetics of this aggregation process and focus on the
long-time asymptotic behavior of the cluster-size density. Our
reference frame is the well-understood behavior in the case of
ordinary aggregation where the cluster-size density is purely
exponential. We find that the density of the smallest clusters is
anomalously large compared with typical-size clusters. This anomaly is
not captured by the scaling function which characterizes the bulk of
the density. We also find an interesting change in the shape of the
size density. In addition to the overpopulation of
smaller-than-typical clusters, there is also an overpopulation of
larger-than-typical clusters. The small-size tail and the large-size
tails are enhanced at the expense of moderate-size clusters. The
enhancement of small clusters is easy to appreciate as it is a direct
consequence of choosing the larger cluster.  The enhancement of large
clusters is an indirect, perhaps counter-intuitive, consequence of the
aggregation rules.

We also study a few other implementations of choice. First,
we consider the case where the smaller candidate cluster participates in the
merging event. In this case, we observe an opposite change in
the shape of the size density. Now, both the small-size tail and the
large-size tail of the size density are suppressed, while the density
of moderate-size clusters is enhanced. Second, we study aggregation
processes where multiple candidates clusters are drawn and the maximal
(or minimal) merges with the target cluster. In the maximal choice
case, we find an interesting sequence of distinct scaling laws
corresponding to the densities of the smallest clusters. Finally, we
also investigate a symmetric implementation of choice where two
candidate {\rm pairs} are drawn at random and one of the pairs
undergoes aggregation. We find that the changes in the shape of the
size density, described above, are generic.

This paper is organized as follows. First, we briefly review ordinary
aggregation, the most basic process where the merging clusters are
chosen at random (Sec.~\ref{sec:no}).  Next, we introduce the notion
of choice by considering the case where the larger of two
randomly-selected clusters merges with another randomly-selected
cluster. From the rate equation for the cluster-size density, we
obtain the density of the smallest cluster, the small-size tail of the
density, as well as the large-size tail of the density
(Sec.~\ref{sec:Max}). We also detail results of our numerical
simulations to gain insights into the entire size density. We apply
the same theoretical tools to the case where the smaller of the two
candidate clusters undergoes merger (Sec.~\ref{sec:Min}), and to the
case where multiple candidates clusters are drawn
(Sec.~\ref{sec:many}).  In Sect.~\ref{sec:Max}--\ref{sec:many} the
choice is implemented asymmetrically as the target cluster was selected
from the outset.  In Sec.~\ref{sec:symmetric}, we introduce a
symmetric implementation of choice where two pairs of clusters are
chosen and only one of these pairs undergoes aggregation. We conclude
in section VII and provide several technical details in the Appendices.

\section{Ordinary Aggregation}
\label{sec:no}

In ordinary aggregation, two clusters are chosen at random and merge
to form a larger cluster \cite{Ald99,FL03,book}. This basic process
can be generalizes to model polymerization \cite{bk}, condensation
\cite{yo}, chemotaxis \cite{sf}, and random structures \cite{bk1,aal}.
Symbolically, we may represent the merger process as $i,j\to i+j$
where the aggregation rate is independent of cluster mass.  The
elementary aggregation step is repeated indefinitely.  Initially, the
system consists of identical particles whose mass can be set to
unity. We tacitly take the thermodynamic limit, that is, assume that
the initial number of particles is infinite.

Two clusters participate in each aggregation event and the number of
clusters declines by one. Hence, the total cluster density $c(t)$
obeys the rate equation 
\begin{equation} 
\label{c-eq}
\frac{dc}{dt}=-c^2\,.
\end{equation}
Without loss of generality, we set the merging rate unity.
Solving \eqref{c-eq} subject to the initial condition $c(0)=1$ yields
\begin{equation}     
\label{c}
c(t)=(1+t)^{-1}.
\end{equation}
In the long-time limit we have $c\simeq t^{-1}$. (In our notations
$a\sim b$ indicates the ratio $a/b$ approaches a constant when
$t\to\infty$, while $a\simeq b$ indicates that the ratio approaches
unity.)

Let $c_k(t)$ be the density of clusters of mass $k$ at time $t$. This
quantity obeys the master equation
\begin{equation}
\label{ck-eq}
\frac{dc_k}{dt}=\sum_{i+j=k}c_ic_j-2c\, c_k\,.
\end{equation}
By summing \eqref{ck-eq} we can verify that the density $c=\sum_k c_k$
obeys \eqref{c-eq}.  The mass density $M=\sum_k k\,c_k$ is conserved
$dM/dt=0$, as also follows from \eqref{ck-eq}.

We shall consider the mono-disperse initial condition
\begin{equation}
\label{IC}
c_k(0)=\delta_{k,1}\,.
\end{equation}
We note that it suffices to use \eqref{IC}, because the asymptotic
behavior is universal as long as the initial density decays rapidly
with mass. The density of the smallest clusters, monomers, obeys
$dc_1/dt=-2cc_1$, from which $c_1(t)=(1+t)^{-2}$.  The monomer density
decays more rapidly than the overall density, $c_1\simeq t^{-2}$.
Starting from \eqref{IC}, the cluster-size density remains purely
exponential
\begin{equation} 
\label{ck}
c_k(t)=\frac{t^{k-1}}{(1+t)^{k+1}}\,,
\end{equation}
throughout the evolution. 

Using mass conservation and the density
decay \eqref{c} alone, we can deduce the average cluster size
\hbox{$\langle k \rangle = M/c$} or $\langle k\rangle=1+t$.  In the
long time-limit, the size distribution attains the scaling form 
\begin{equation} 
\label{F-def}
c_k(t)\simeq t^{-2}F(kt^{-1})\,.
\end{equation}
This form reflects the linear growth of the typical mass $k\sim
t$. According to the density decay $c\simeq t^{-1}$ and mass
conservation, $M=1$, the scaling function must satisfy two
constraints:
\begin{equation}
\label{F-int}
\int_0^\infty dx\,F(x)=1, \qquad \int_0^\infty dx\,x\, F(x)=1\,.
\end{equation}
For ordinary aggregation Eq.~\eqref{ck} implies that the scaling
function is purely exponential, $F(x)=e^{-x}$, a behavior that holds
for any (rapidly decaying) initial condition.

Ordinary aggregation provides a useful reference point for our study.
Throughout this study the density satisfies \eqref{c-eq}, and mass is
certainly conserved.  Moreover, the size density generally follows the
scaling form \eqref{F-def}, with the scaling function satisfying the
constraints \eqref{F-int}.

\section{Maximal Choice}
\label{sec:Max}

We now incorporate choice while preserving most features of ordinary
aggregation. In particular, aggregation remains a {\em binary} process
with two clusters joining to form one larger cluster
(Fig.~\ref{fig-max}). One cluster with size $i$ is selected at random,
and it is certain to participate in the aggregation process. The
aggregation partner is selected as the larger of two, randomly
selected clusters of sizes $j_1$ and $j_2$. Schematically,
\begin{equation}
\label{process-max}
i,j_1,j_2\to i+{\rm max}(j_1,j_2), ~{\rm min}(j_1,j_2)\,.
\end{equation}
We reiterate that while three clusters are drawn, only two 
undergo aggregation. Mass is of course conserved and we consider 
the mono-disperse initial condition \eqref{IC}. 

As in ordinary aggregation, two clusters are lost in each aggregation
event and one new cluster is formed. Hence, the total density obeys
\eqref{c-eq}, and it decays according to \eqref{c}.
Consequently, the growth of the typical mass as well as the scaling
form \eqref{F-def} with the constraints \eqref{F-int} hold.

The cluster-size density obeys the master equation 
\begin{eqnarray}
\label{c-eq-max}
\frac{dc_k}{dt} = c^{-1}\!\!\sum_{i+j=k}c_i\left(g_j^2-g_{j-1}^2\right)
-c\,c_k-\left(g_k^2-g_{k-1}^2\right).
\end{eqnarray}
Here, $g_k=\sum_{l\leq k}c_l$ is the cumulative size density, namely,
the density of clusters with size smaller than or equal to $k$.  The
gain term has the same convolution structure as \eqref{ck-eq} with one
density corresponding to the target cluster and another density
corresponding to the larger of the two candidate clusters. The
quantity $g_k^2-g_{k-1}^2$ is proportional to the probability that the
largest of two randomly selected clusters has size $k$, and the
multiplicative constant $c^{-1}$ ensures proper normalization.  There
are two loss terms. The first represents the target cluster, and the
second accounts for the selected cluster. One can verify that the
total cluster density obeys \eqref{c-eq}.

Throughout this study, we repeatedly make the transformations 
\begin{equation}
\label{transform}
C_k = \frac{c_k}{c} \qquad \text{and}\qquad \tau = \ln (1+t). 
\end{equation}
The distribution $C_k$ is normalized, \hbox{$\sum_k
  C_k=1$}, and it represents the fraction of clusters of size $k$. It
is also convenient to introduce the time variable $\tau$ which
satisfies \hbox{$dc/dt=1/c$}.  With the transformations
\eqref{transform}, the first loss term in \eqref{c-eq-max} is
eliminated, and we arrive at 
\begin{eqnarray}
\label{ck-eq-max}
\frac{dC_k}{d\tau} = \sum_{i+j=k}C_i\left(G_j^2-G_{j-1}^2\right)
-\left(G_k^2-G_{k-1}^2\right)\,.
\end{eqnarray}
Here $G_k=\sum_{l\leq k}C_l$ is the cumulative size distribution, the
fraction of clusters with size not exceeding $k$.

For monomers, $k=1$, we have $dC_1/d\tau=-C_1^2$ and since $C_1(0)=1$,
then $C_1(\tau)=(1+\tau)^{-1}$. In terms of the actual time variable, the
density of monomers reads
\begin{equation}
\label{c1-max}
c_1(t)=[(1+t) + (1+t)\ln (1+t)]^{-1}\,.
\end{equation}
The asymptotic behavior $c_1\simeq (t\ln t)^{-1}$ represents a
substantial enhancement over the monomer density $c_1\simeq t^{-2}$
for ordinary aggregation. As expected, monomers become populous because
they are  least likely to participate in the aggregation process
\eqref{process-max}.

A more elaborate calculation  (see Appendix A) gives the density of dimers: 
\begin{equation}
\label{c2-max}
c_2=e^{-\tau}u^3\,\frac{I_0(2)K_0(2\,u)-K_0(2)I_0(2\,u)}{I_0(2)K_1(2\,u)+K_0(2)I_1(2\,u)}\,.
\end{equation}
Here $I_\nu$ and $K_\nu$ are the modified Bessel functions with index
$\nu$, and $u=(1+\tau)^{-1/2}$. 
Using the asymptotic relations,
$K_0(2u)\simeq \ln(1/u)$ and $K_1(2u)\simeq (2u)^{-1}$
when $u\to 0$, we find the asymptotic decay 
\begin{equation}
\label{c2-asymp-max}
c_2(t)\simeq \frac{1}{t}\,\frac{\ln(\ln t)}{(\ln t)^2}\,.
\end{equation}
The dimer density is much smaller the monomer density,
$\frac{c_2}{c_1}\simeq \frac{\ln(\ln t)}{\ln t}$ when $t\gg 1$. In
comparison with ordinary aggregation where $c_2\simeq t^{-2}$, the
dimer density \eqref{c2-asymp-max} is substantially larger, however.

For trimers and other finite clusters, $k\geq 3$, we can obtain the
leading asymptotic behavior.  As for monomers and dimers, the loss
rate in \eqref{ck-eq-max} dominates. Furthermore, since $C_1\gg C_2$
in the asymptotic regime, the dominant term in Eq.~\eqref{ck-eq-max}
involves the monomer fraction,
\begin{equation}
\label{ck-small-max}
\frac{dC_k}{d\tau}\simeq - 2C_1C_k\,.
\end{equation}
We now substitute the asymptotic behavior $C_1\simeq \tau^{-1}$,  
and immediately obtain $C_k\sim \tau^{-2}$. In terms of the physical
time variable 
\begin{equation}
\label{ck-max}
c_k(t)\sim \frac{1}{t}\,\frac{1}{(\ln t)^2}
\end{equation}
for $3\leq k \ll t$. Hence, $c_1\gg c_2\gg c_k$ when $k\geq 3$. 
The ratio $\frac{c_2}{c_3}\sim \ln(\ln t)$ diverges with time, but very slowly. 

In summary, equations \eqref{c1-max}, \eqref{c2-max}, and
\eqref{ck-max} show that there are three distinct scaling laws 
for small clusters
\begin{equation}
\label{ck-asymp-max}
c_k\sim 
\begin{cases}
\frac{1}{t}\,\frac{1}{(\ln t)}                     & k=1 \cr 
\frac{1}{t}\,\frac{\ln(\ln t)}{(\ln t)^2}        & k=2 \cr
\frac{1}{t}\,\frac{1}{(\ln t)^2}                  & k\geq 3 .\cr
\end{cases} 
\end{equation}
As a consequence of choice, there is a strong enhancement of small
clusters compared to ordinary aggregation. Further, three different
decay laws characterize the density of monomers, dimers, and clusters
of mass $3\leq k \ll t$. As we show below, the scaling function
underlying the cluster-size density captures $c_k$ with $k\geq 3$.

\begin{figure}[t]
\includegraphics[width=0.44\textwidth]{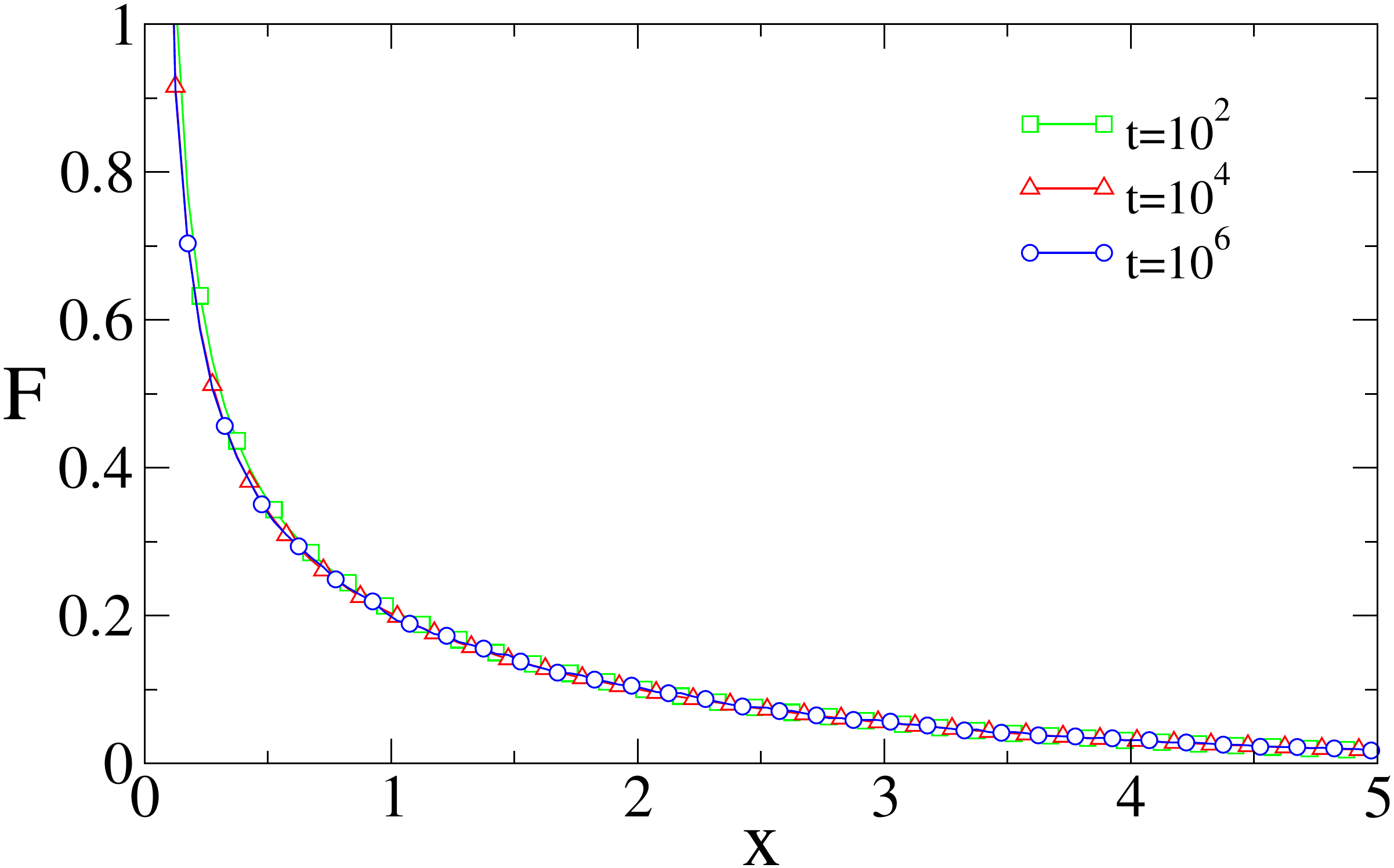}
\caption{The scaling function $F(x)$ in the maximal choice model. Shown is 
  $F(x) \equiv t^2 c_k(t)$ versus the scaling variable $x=k/t$ at three
  different times.}
\label{fig-fx-max}
\end{figure}

Our numerical simulations (see Fig.~\ref{fig-fx-max}) confirm
that the cluster-size density adheres to the scaling form
\eqref{F-def}: in terms of the properly normalized cluster size
$x=k/t$, the size density has a universal shape in the asymptotic
regime. The scaling function $F(x)$ satisfies the integro-differential
equation
\begin{eqnarray}
\label{Fx-eq-max}
\frac{d[xF(x)]}{dx}+\int_0^x dy F(x-y)\frac{d\Phi^2(y)}{dy}-
\frac{d\Phi^2(x)}{dx}=0\,.
\end{eqnarray}
Here $\Phi(x)=\int_0^x dy F(y)$ is the fraction of clusters with size
smaller than $x=k/t$ in the long-time limit.  To obtain
\eqref{Fx-eq-max} we simply substitute \eqref{F-def} into the rate
equation \eqref{ck-eq-max}. The two nonlinear terms in
\eqref{Fx-eq-max} correspond to the two nonlinear terms in
\eqref{ck-eq-max}.

First, we consider statistics of small clusters. As mentioned
above, the convolution term, which corresponds to generation of larger
clusters from smaller clusters through aggregation, is
negligible. Keeping only the leading terms when $x\ll 1$, we get
\begin{eqnarray}
\label{Fx-eq-smallx-max}
\frac{d}{dx}\left[x\,F-\Phi^2(x)\right]=0\,.
\end{eqnarray}
Hence $xF=\Phi^2$, or alternatively
$x\Phi'=\Phi^2$.  Solving this differential equation yields
\hbox{$\Phi=[\ln(1/x)]^{-1}$}  leading to the asymptotic
behavior
\begin{equation}
\label{Fx-smallx-max}
F(x)\simeq \frac{1}{x}\,\frac{1}{[\ln(1/x)]^2}\,
\end{equation}
as $x\to 0$.  This form is consistent with the cluster density
\eqref{ck-max}, and it specifies the proportionality constant:
\hbox{$c_k\simeq k^{-1}\,[t\,(\ln t)^2]^{-1}$}. The diverging
small-$x$ tail of the scaling function does not capture the anomalous
populations of monomers and dimers (see Fig.~\ref{fig-fx-max}).

\begin{figure}[t]
\includegraphics[width=0.44\textwidth]{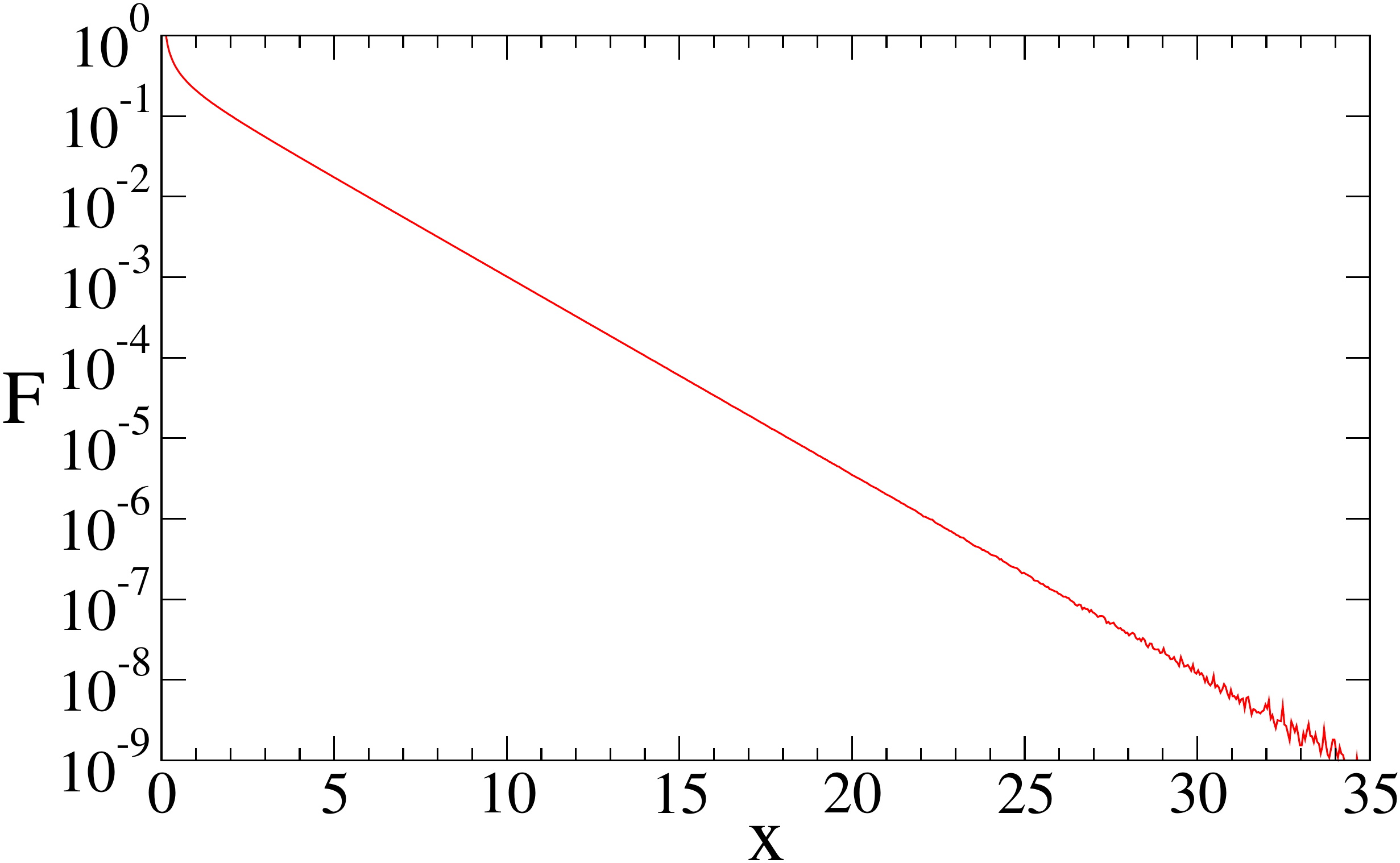}
\caption{The large-size tail of the scaling function $F(x)$ showing 
  the exponential decay \eqref{Fx-largex-max}.}
\label{fig-fx-tail-max}
\end{figure}

Next, let us consider statistics of large clusters. In the limit $x\gg
1$, the convolution term is dominant, and the governing equation
\eqref{Fx-eq-max} becomes
\begin{equation}
\label{Fx-max-large}
xF'(x)+2\int_0^x dy F(y)F(x-y)=0\,.
\end{equation}
Here, we also assumed $xF'\gg F$ which can be justified a
posteriori. Equation \eqref{Fx-max-large} is essentially the same as
in ordinary aggregation, and therefore, the tail is exponential:
\begin{equation}
\label{Fx-largex-max}
F(x)\simeq \frac{\alpha}{2}e^{-\alpha x}
\end{equation}
when $x\to \infty$.  Our numerical simulations confirm this
exponential asymptotic decay (see Fig.~\ref{fig-fx-tail-max}) with the
constant $\alpha=0.57\pm 0.01$. 

The tail \eqref{Fx-smallx-max} shows that the density of
small clusters is enhanced compared with ordinary aggregation: $F(x)
\gg  e^{-x}$ when $x\to 0$. This is an expected
consequence of choice --- very small clusters are less likely to
participate in aggregation, so their population is
enhanced. Remarkably, the same holds for large
clusters --- since $\alpha<1$, the large-size tail 
\eqref{Fx-largex-max} is enhanced compared with ordinary aggregation,
$F(x)\gg e^{-x}$ when $x\to\infty$. This is an indirect
consequence of choice --- somehow very large clusters are ``shielded''
from aggregation.

\begin{figure}[t]
\includegraphics[width=0.44\textwidth]{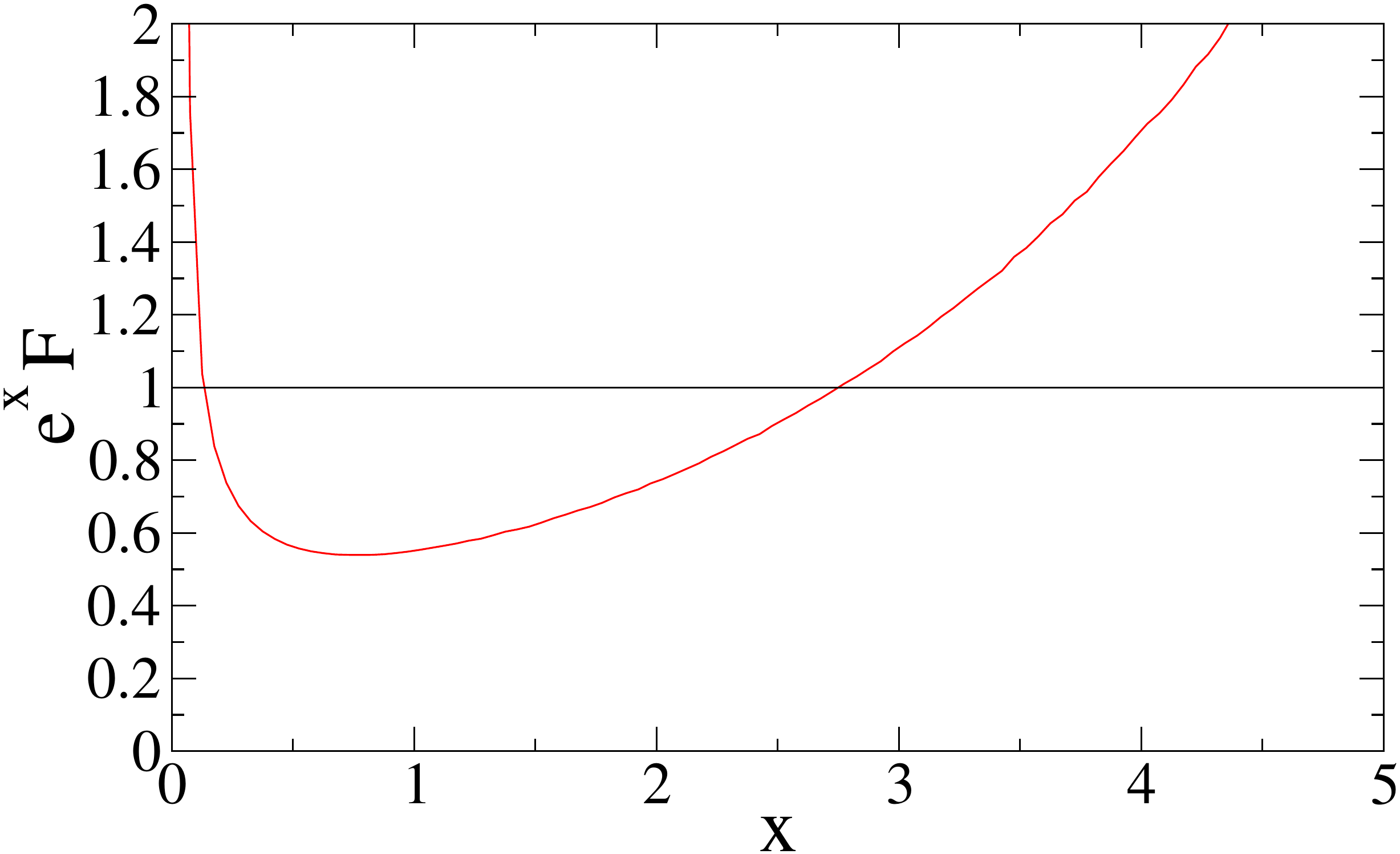}
\caption{The normalized scaling function $e^x\,F(x)$ versus the
  scaling variable $x$.  Also shown for reference is the unit constant
  corresponding to the ordinary aggregation case.}
\label{fig-fx-max-compare}
\end{figure}

Figure \eqref{fig-fx-max-compare} compares aggregation with choice
with ordinary aggregation, and it demonstrates that there are three
regimes of behavior, as the normalized scaling function $e^{x}F(x)$ is
non-monotonic.  Small clusters with size $x<x_1$ are overpopulated 
compared with ordinary aggregation. Large clusters with size $x>x_2$
are also overpopulated compared with ordinary aggregation. The
conservation laws \eqref{F-int} dictate that clusters of moderate
sizes $x_1<x<x_2$ must be underpopulated. Hence, introducing choice
alters the shape of the size density. 

Monte Carlo simulations of aggregation processes are rather 
straightforward when the aggregation rate is uniform as is the
case of the merging rule \eqref{process-max} and other rules studied in
this paper.  Initially, the system consists of $N_0$ identical
particles with unit mass. In each aggregation event, three {\em
  distinct} particles are selected at random. One of these particles
is designated as the target particle, and it merges with the larger of
the remaining two particles. When $N_0$ is large, the overall density
\eqref{c} specifies time as $t\equiv N_0/N$ where $N$ is the number of
remaining aggregates. The simulation results presented throughout this
paper were obtained using $N_0=10^8$, and an average over roughly 
$10^5$ independent realizations.

\section{Minimal Choice} 
\label{sec:Min}

We now consider the complementary case where the target cluster merges
with the smaller of the two candidate clusters (see also Fig.~\ref{fig-max}) 
according to the scheme 
\begin{equation}
\label{process-min}
i,j_1,j_2\to i+{\rm min}(j_1,j_2),~{\rm max}(j_1,j_2)\,.
\end{equation}
As in maximal choice mass is conserved, and the total density decays
according to \eqref{c}.

The size-density $c_k(t)$ satisfies the master equation 
\begin{eqnarray}
\label{c-eq-min}
\frac{dc_k}{dt} =c^{-1} \!\!\!\sum_{i+j=k}c_i\left(h_{j}^2-h_{j+1}^2\right)
-c\,c_k-\left(h_{k}^2-h_{k+1}^2\right)
\end{eqnarray}
subject to \eqref{IC}.  The quantity $h_k=\sum_{l\geq k} c_l$ is the
density of clusters of size larger than or equal to $k$. The
cumulative distributions $h_k$ and $g_{k-1}$ appearing in
\eqref{c-eq-max} are complimentary: $g_{k-1}+h_{k}=c$ for all $k\geq
1$.  As in Eq.~\eqref{c-eq-max}, the first loss term in
Eq.~\eqref{c-eq-min} corresponds to the target cluster and the second,
to the selected cluster. The quantity $h_{k}^2-h_{k+1}^2$ is
proportional to the probability that the selected cluster has size
$k$. By summing \eqref{c-eq-min}, we can verify that the density
satisfies \eqref{c-eq}.

In terms of the modified time variable $\tau$, the size distribution
$C_k$ satisfies
\begin{eqnarray}
\label{ck-eq-min}
\frac{dC_k}{d\tau} = \sum_{i+j=k}C_i\left(H_{j}^2-H_{j+1}^2\right)
-\left(H_{k}^2-H_{k+1}^2\right). 
\end{eqnarray}
Here $H_k=\sum_{l\geq k}C_l$. We note that $G_k+H_{k+1}=1$ and
$H_1=1$ at all times.  The initial condition \eqref{IC} becomes
$C_k(0)=\delta_{k,1}$. 

According to \eqref{ck-eq-min} the density of monomers satisfies
$\frac{dC_1}{d\tau}=C_1^2-2C_1$, with $C_1(0)=1$.  This Bernoulli
equation is solved to yield
\hbox{$C_1(\tau)=2/(1+e^{2\tau})$}. In terms of the original time
variable, the density of monomers reads
\begin{equation}
\label{c1-min}
c_1(t)=\frac{2}{(1+t)+(1+t)^3}\,.
\end{equation}
In the long-time limit we have $c_1(t)\simeq 2\,t^{-3}$, whereas in
ordinary aggregation $c_1(t)\simeq t^{-2}$. Monomers are most likely
to participate in the aggregation process \eqref{process-min}, and
consequently, they decay rapidly.

\begin{figure}[t]
\includegraphics[width=0.44\textwidth]{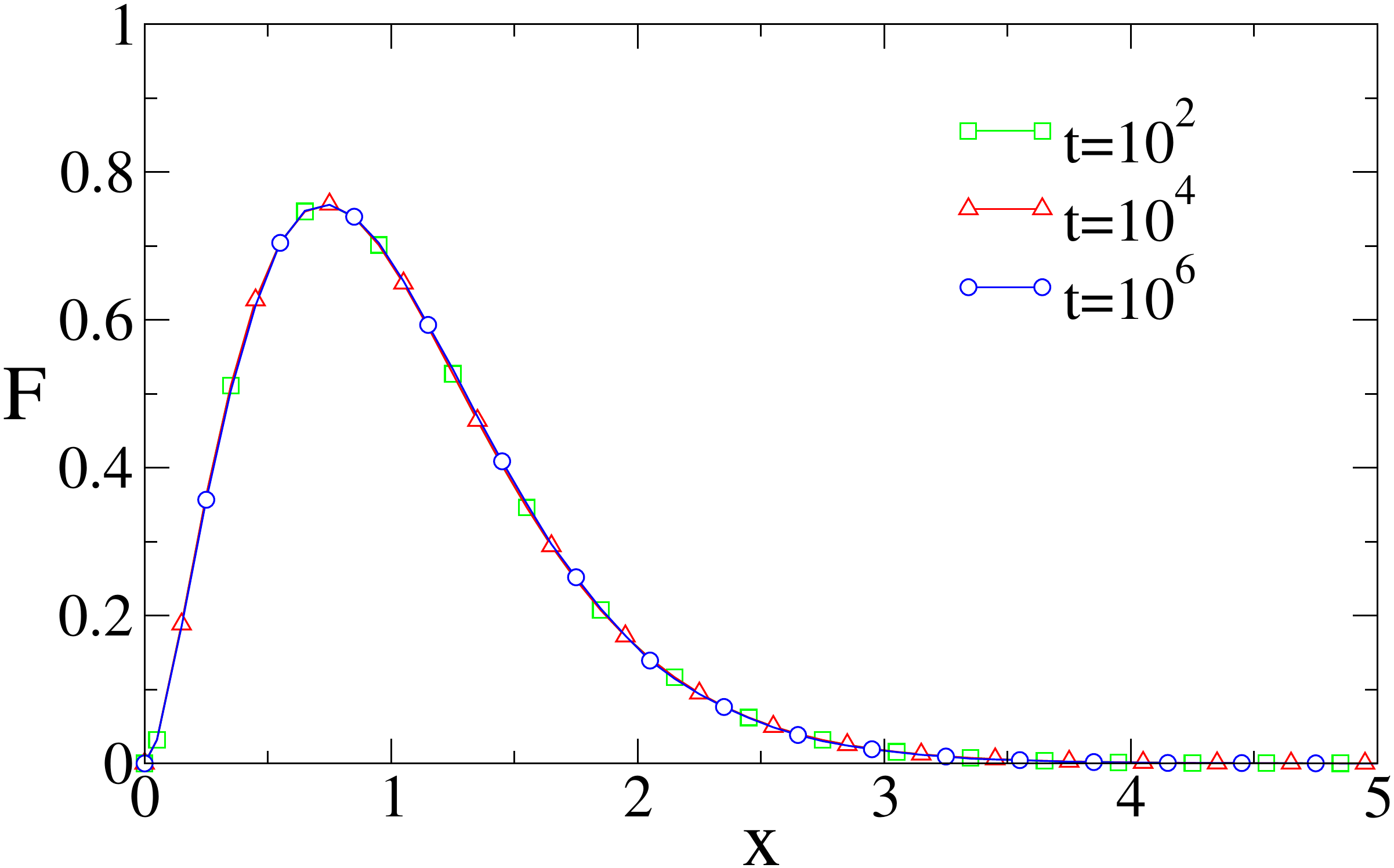}
\caption{The scaling function $F(x)$ in the minimal choice model.
  Shown is $F(x)\equiv t^2 c_k(t)$ versus the scaling variable $x=k/t$
  at three different times.}
\label{fig-fx-min}
\end{figure}

One can also obtain the exact expression for the dimer density (see
Appendix A)
\begin{equation}
\label{c2-min}
c_2 = \frac{(e^{-\tau}\,v)^3}{8}\,
\frac{J_0(2)Y_0(v)-Y_0(2)J_0(v)}{Y_0(2)J_1(v)-J_0(2)Y_1(v)}\,.
\end{equation}
Here $J_\nu$ and $Y_\nu$ are the Bessel functions with index $\nu$ and
$v =[8/(1+e^{-2\tau})]^{-1/2}$.  Asymptotically, the dimer density
decays according to $c_2(t)\simeq A_2\,t^{-3}$ with the 
prefactor
\begin{eqnarray}
\label{ratio}
A_2=\sqrt{8}\,\,
\frac
{J_0(2)Y_0(\sqrt{8})\!-\!Y_0(2)J_0(\sqrt{8})}
{Y_0(2)J_1(\sqrt{8})\!-\!J_0(2)Y_1(\sqrt{8})}\,
\end{eqnarray}
or $A_2=3.878012\cdots$. In contrast with the behavior
\eqref{ck-asymp-max}, the ratio $c_2/c_1$ now approaches a nontrivial
constant.

For finite but small $k$, the loss rate in \eqref{ck-eq-max}
dominates. By using $H_k-H_{k+1}=C_k$ we have $dC_k/d\tau\simeq
-2C_k$, and therefore $C_k(\tau)\sim e^{-2\tau}$. In general, the
density of small clusters decays algebraically,
\begin{equation}
\label{ck-asymp-min}
c_k(t)\simeq A_k\,t^{-3}\,,
\end{equation}
for finite $k\ll t$. As expected, small clusters are suppressed due to
choice. In contrast with maximal choice, however, there are no
anomalies associated with monomers or with dimers, and a single
scaling law characterizes small clusters.  As shown below, the decay
\eqref{ck-asymp-min} is captured by the scaling function $F(x)$.

Our numerical simulations confirm that once size is rescaled by the
typical size, $k\simeq t$, the size distribution becomes universal in
the long-time limit (see Fig.~\ref{fig-fx-min}). By substituting the
scaling ansatz \eqref{F-def} into the governing equation
\eqref{c-eq-min}, we find that the scaling function obeys
\begin{eqnarray}
\label{Fx-eq-min}
\frac{d[xF(x)]}{dx} - \int_0^x dy F(x-y)\frac{d\Psi^2(y)}{dy} 
+ \frac{d\Psi^2(x)}{dx}=0\,.
\end{eqnarray}
Here $\Psi(x)=\int_x^\infty dy F(y)$ is the fraction of clusters of
size larger than $x=k/t$. Once again, the scaling function obeys the
two constraints in \eqref{F-int}.

\begin{figure}[t]
\includegraphics[width=0.44\textwidth]{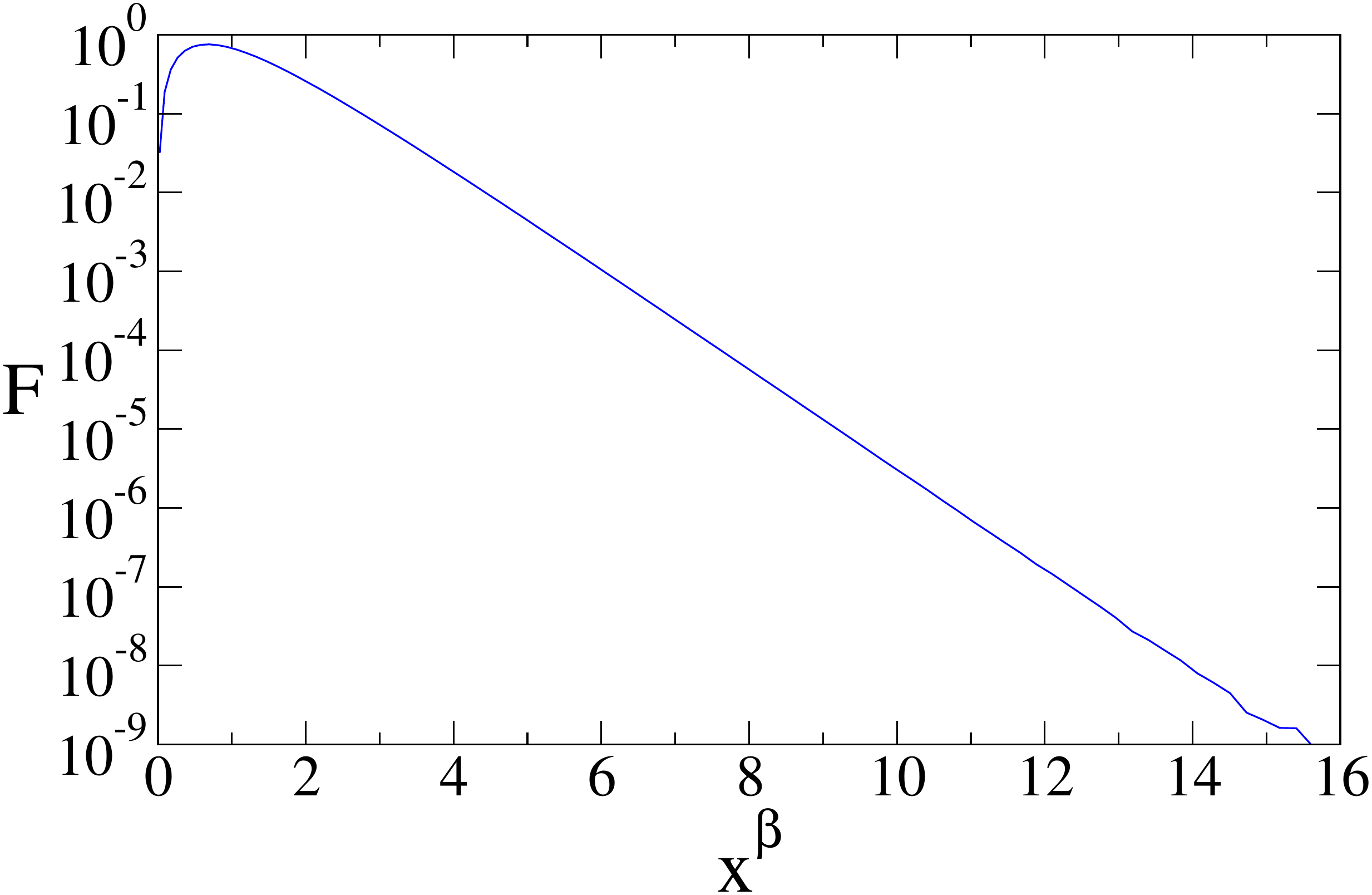}
\caption{The large-$x$ tail of the scaling function $F(x)$ showing the
  super-exponential decay \eqref{Fx-largex-min} with $\beta\cong
  1.26749$.}
\label{fig-fx-tail-min}
\end{figure}

First, we discuss statistics of small sizes. The convolution term is
negligible when $x\ll 1$ and using \hbox{$\Psi'(x)=-F(x)$} we get
$xF'(x)=F(x)$. Therefore the scaling function is linear (see
Fig.~\ref{fig-fx-min})
\begin{eqnarray}
\label{Fx-smallx-min}
F(x)\sim x\,,
\end{eqnarray}
in the limit $x\to 0$. The linear behavior confirms
\eqref{ck-asymp-min}, and further, it shows that $A_k\sim k$ for large
but finite $k$.  In contrast with maximal choice, the equation
governing $F(x)$ is linear in the limit $x\to 0$, and determining the
proportionality constant in $F(x)\simeq {\rm const.}\times x$ requires
a full solution of the nonlinear equation \eqref{Fx-eq-min}.

Let us now consider the large-$x$ behavior. Since the convolution term
is dominant in \eqref{Fx-eq-min}, we have 
\begin{equation}
\label{Fx-eq-largex-min}
\frac{d[xF(x)]}{dx}+ 2\int_0^x dy F(x-y)\frac{d\Psi^2(y)}{dy}=0
\end{equation}
when $x\gg 1$. We anticipate (and justify a posteriori) a sharp decay of the scaling function.
In this scenario, \hbox{$\Psi(x)=\int_y^\infty dz
  F(z)\asymp F(y)$} and $-F'(x)\asymp F(x)$. (We use the notation
$A\asymp B$ to imply that the logarithms of $A$ and $B$ have the same
asymptotic behavior, $\ln A \simeq \ln B$.) Further, we postulate that
the integral in \eqref{Fx-eq-largex-min} is maximal at $y=\sigma\,x$,
with $0<\sigma<1$, and therefore
\begin{equation}
\label{F-balance}   
F(x)\asymp F^2(\sigma\,x)F(x-\sigma\,x)\,.
\end{equation}
Taking the logarithm of both sides we arrive at a {\em linear}
functional equation $\ln F(x)=2\ln F(\sigma\,x) + \ln
F(x-\sigma\,x)$\,.  This equation admits a simple family of algebraic
solutions, $\ln F(x)\simeq -{\rm const}\times x^\beta$, or
equivalently,
\begin{equation}
\label{Fx-largex-min}
F(x)\asymp \exp\left(-{\rm const.}\times x^\beta\right)\,
\end{equation}
with exponent $\beta>1$. The exponent $\beta$ and the parameter $\sigma$ 
are related via 
\begin{equation}
\label{sigma}
2\sigma^\beta+(1-\sigma)^\beta=1\,.
\end{equation}
An additional relation is needed to ``select'' $\beta$.  Selection
problems arise in the context of nonlinear
partial differential equations \cite{van03} and nonlinear recurrences
\cite{MK03}. Typically, the selection criterion is tied to an
extremum, as is the case for velocity selection in traveling waves
\cite{van03}. Guided by these examples, we postulate that $\beta$ is
selected by the requirement that the quantity $\sigma\equiv
\sigma(\beta)$, determined by Eq.~\eqref{sigma}, increases with {\em
  maximal} rate at the selected $\beta$, that is $d\sigma/d\beta$ is
maximal. This extremum requirement specifies the selection criterion
\begin{equation}
\label{criterion}
\frac{d^2\sigma}{d\beta^2}=0\,.
\end{equation}
Using equations \eqref{sigma} and \eqref{criterion} we obtain (see
Appendix B for further details) $\beta\cong 1.26749$ and the
corresponding $\sigma \cong 0.166453$. Our simulation results support
this value, as shown in figure \ref{fig-fx-tail-min}. The
super-exponential tail for large $x$ is sufficiently sharp to provide
an a posteriori justification to the assumptions made in deriving
\eqref{sigma}. 

\begin{figure}[t]
\includegraphics[width=0.44\textwidth]{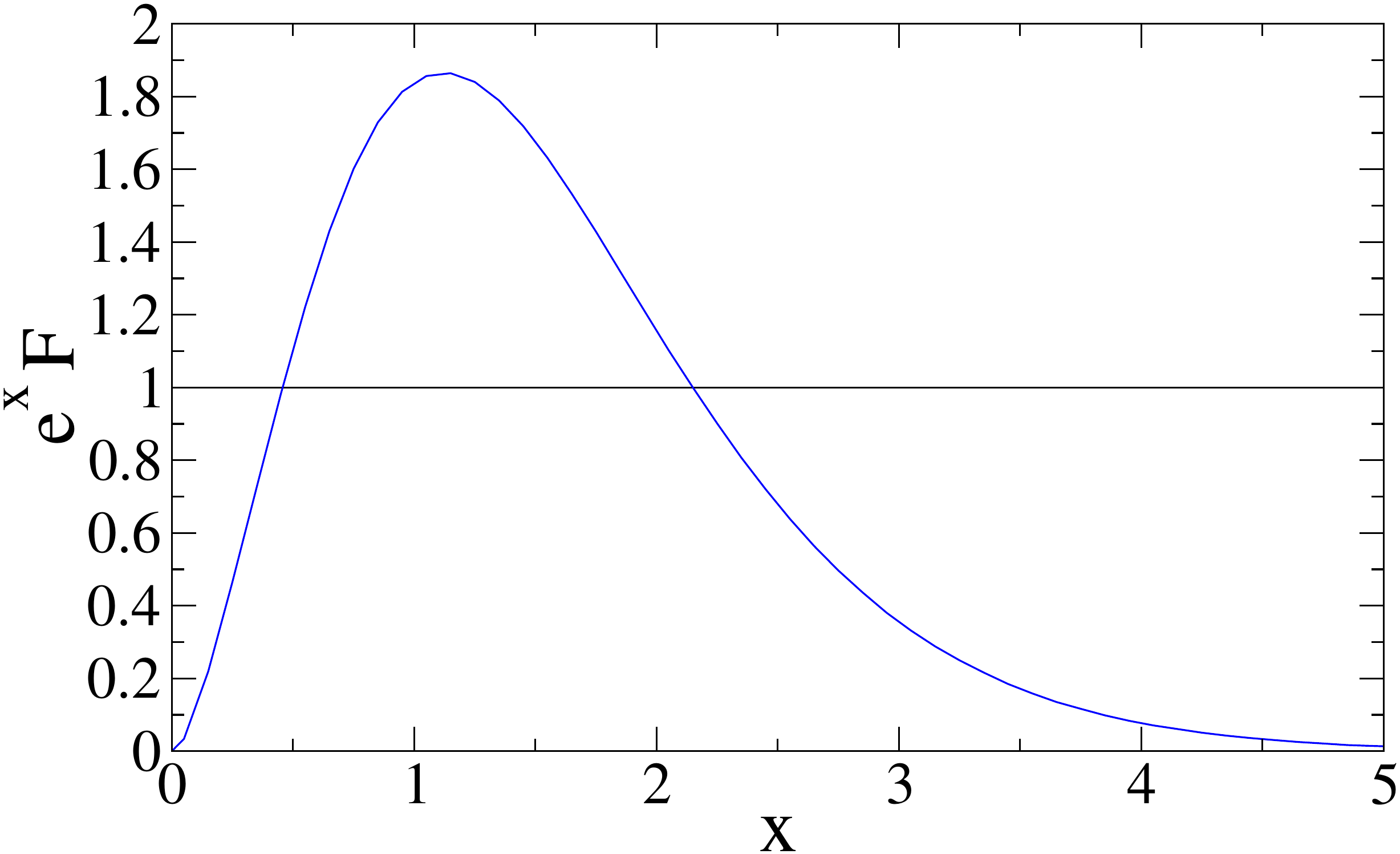}
\caption{The normalized scaling function $e^x\,F(x)$ versus the
  scaling variable $x$.  Also shown for reference is the unit
  constant, corresponding to the ordinary aggregation case.}
\label{fig-fx-min-compare}
\end{figure}

The small-size tail \eqref{Fx-smallx-min} confirms that when the
smaller of the two candidate clusters undergoes aggregation, the
population of small clusters is suppressed.  The large-size tail
\eqref{Fx-largex-min} is much steeper than exponential: $F(x)\ll
e^{-x}$ for large $x$.  Figure \ref{fig-fx-min-compare} compares the
scaling function for minimal choice with ordinary aggregation. There
are three regimes of behaviors: in the small mass range $x<x_1$ and in
the large mass range $x>x_2$ the cluster-size density is
underpopulated while in the intermediate size range $x_1<x<x_2$ the
density is overpopulated. Hence, the effect on size density is the
exact opposite of that found for the maximal case.

\section{Multiple Choice}
\label{sec:many}

In Sections \ref{sec:Max} and \ref{sec:Min} we showed that choice
between two alternatives significantly affects the size density. What
happen if we allow choice between more than two alternatives? In the context 
of other models \cite{broder,adler,michael}, the
general conclusion was that multiple choice modifies the
behavior only quantitatively. As we show below, introduction of
multiple choice in aggregation has interesting consequences, including 
some quantitative changes. 

\subsection{Maximal Choice}

We start with the maximal case, and introduce multiple choice as to
preserves the binary nature of the aggregation process. As in section
\ref{sec:Max}, we choose a single target cluster along with $n$
candidate clusters. The target cluster merges with the largest of
these $n$ clusters, while the rest of the $n-1$ clusters are not
affected. This merger process preserves the total mass, and the total
cluster density is given by \eqref{c}.

The cluster-size density $c_k(t)$ obeys 
\begin{eqnarray}
\label{c-eq-mmax}
\frac{dc_k}{dt} = \sum_{i+j=k}c_i\frac{g_j^n-g_{j-1}^n}{c^{n-1}}
-c\,c_k-\frac{g_k^n-g_{k-1}^n}{c^{n-2}}\,,
\end{eqnarray}
where $g_k=\sum_{l\leq k}c_l$ is the cumulative density. The master
equation \eqref{c-eq-mmax} reduces to \eqref{ck-eq} and
\eqref{ck-eq-max} when $n=1$ and $n=2$, respectively.  The quantity
$g_k^n-g_{k-1}^n$ is proportional to the probability that the selected
cluster has size $k$.  From \eqref{c-eq-mmax} we can obtain the master
equation governing the normalized cluster-size distribution $C_k =
c_k/c$. Using the time variable $\tau=\ln (1+t)$ and $G_k=\sum_{l\leq
  k} C_l$ we get
\begin{eqnarray}
\label{ck-eq-mmax}
\frac{dC_k}{d\tau} = \sum_{i+j=k}C_i\left(G_{j}^n-G_{j-1}^n\right)
-\left(G_{k}^n-G_{k-1}^n\right).
\end{eqnarray}
The density of monomers satisfies $dC_1/d\tau=-C_1^n$, from which
\hbox{$C_1(\tau) = [1+(n-1)\tau]^{-1/(n-1)}$}\,. In terms of the physical 
time
\begin{equation}
\label{c1-mmax}
c_1(t)=(1+t)^{-1}\left[1+(n-1)\ln(1+t)\right]^{-1/(n-1)}\,.
\end{equation}
This decay represents an enhancement over ordinary aggregation.

For finite cluster size $k$, it is possible to proceed with asymptotic
analysis of \eqref{ck-eq-max} and find $C_k\sim C_1^k$ for $k<n$ and
$C_k\sim C_1^n$ for $k>n$. The three-tier asymptotic behavior
\eqref{ck-asymp-max} generalizes as follows 
\begin{equation}
\label{ck-asymp-mmax}
c_k(t)\sim 
\begin{cases}
\frac{1}{t}\,\frac{1}{(\ln t)^{k/(n-1)}} & k<n\,,\\
\frac{1}{t}\,\frac{\ln(\ln t)}{(\ln t)^{n/(n-1)}} & k=n\,,\\
\frac{1}{t}\,\frac{1}{(\ln t)^{n/(n-1)}} & k>n\,.
\end{cases} 
\end{equation}
Interestingly, there are $n$ distinct scaling laws that characterize
the enhancement of small clusters. The density of monomers is the
largest, the density of dimers is the next largest and so on.  Thus,
multiple choice leads to multiple anomalies in the cluster-size
density.
 
The scaling function now obeys an 
integro-differential equation 
\begin{eqnarray}
\label{Fx-eq-mmax}
\frac{d}{dx}\left[x\,F-\Phi^n(x)\right]+\int_0^x dy\, F(x-y) 
\frac{d\Phi^n(y)}{dy}=0
\end{eqnarray}
which generalizes \eqref{Fx-eq-max}. Here $\Phi(x)=\int_0^x dy F(y)$
is the fraction of clusters with size smaller than $x$. The tails of
the scaling function are derived by repeating the steps leading to
\eqref{Fx-smallx-max} and \eqref{Fx-largex-max} to give
\begin{eqnarray}
\label{Fx-tails-mmax}
F(x) \simeq 
\begin{cases}
\frac{1}{x}\,\frac{1}{[(n-1)\ln(1/x)]^{n/(n-1)}}  & x\ll 1\, ,\cr 
\frac{\alpha}{n}\exp(-\alpha\,x)      & x\gg 1\,.
\end{cases} 
\end{eqnarray}
The small-$x$ tail captures the behavior of clusters with size $n<k\ll
t$, and the logarithmic divergence reflects the relative abundance
of small clusters due to choice. The divergence in the limit $x\to 0$
becomes weaker and weaker as $n$ grows.  Based on the behavior in the
case $n=2$ we anticipate that $\alpha<1$ in general, and that there is
also an increase in the density of large clusters compared with
ordinary aggregation.

\subsection{Minimal choice}

In the complementary case of minimal choice, the target cluster merges
with the smallest of $n$ candidate clusters. In terms of the modified
time variable $\tau$, the cluster-size distribution $C_k$ satisfies
\begin{eqnarray}
\label{ck-eq-mmin}
\frac{dC_k}{d\tau} = \sum_{i+j=k}C_i\left(H_{j}^n-H_{j+1}^n\right)
-\left(H_{k}^n-H_{k+1}^n\right),
\end{eqnarray}
with $H_k=\sum_{l\geq k}C_l$.  The master equation \eqref{ck-eq-mmin}
generalizes \eqref{ck-eq-min} which corresponds to the case $n=2$.

For finite and small $k$, the leading asymptotic behavior is purely
algebraic as in \eqref{ck-asymp-min}
\begin{equation}
\label{ck-asymp-mmin}
c_k(t)\simeq A_k t^{-n-1}\,.
\end{equation}
This behavior readily follows from \eqref{ck-eq-mmin} by noting that
the dominant term is linear, that is, $dC_k/d\tau\simeq -nC_k$. Hence,
$C_k\sim e^{-n\tau}$ and \eqref{ck-asymp-mmin} follows. The
small-cluster densities \eqref{ck-asymp-mmin} confirm that small
clusters are suppressed when the minimal cluster is chosen for
aggregation.

For monomers, it is possible to obtain the constant $A_1$
analytically.  The monomer concentration obeys
$dC_1/d\tau=(1-C_1)^n-1$, from which 
\begin{equation}
\label{c1-int-mmin}
\int_{C_1}^1 \frac{dv}{1-(1-v)^n}=\tau\,.
\end{equation}
One can evaluate this integral in the asymptotic limit where the lower
limit of the integration vanishes to confirm the decay
\eqref{ck-asymp-mmin}. Moreover, the general expression for the
amplitude is
\begin{equation}
\label{A1-mmin}
A_1 = \exp\left\{ \int_0^1 dv\left[\frac{n}{1-(1-v)^n}-\frac{1}{v}\right] \right\}\,.
\end{equation}
The amplitudes $A_1$ for $n\leq 6$ are listed in Appendix C.

\begin{table}
\begin{tabular}{| c | c | c | c | c | c |}
\hline
$n$ & 2 & 20 & 200 & 2000 & 20000  \\ 
\hline
$\beta$ & 1.26749 & 2.14474 & 3.05326 & 3.99381 & 4.9607 \\ 
\hline
\end{tabular}
\caption{The exponent $\beta$ obtained by solving \eqref{sigma-mmin} and
  \eqref{crit} for $n=2, 20, 200, 2000, 20000$.}
\label{Tab:bba}
\end{table} 

The scaled mass distribution function $F(x)$ satisfies the general
version of \eqref{Fx-eq-min},
\begin{eqnarray}
\label{F-eq-mmin}
\frac{d[xF(x)]}{dx}- \int_0^x dy F(x-y)\frac{d\Psi^n(y)}{dy} 
+ \frac{d\Psi^n(x)}{dx}=0\,.
\end{eqnarray}
Here, $\Psi(x)=\int_x^\infty dy F(y)$. By repeating the steps leading
to the tails \eqref{Fx-smallx-min} and \eqref{Fx-largex-min}, we obtain
the leading asymptotic behaviors
\begin{eqnarray}
\label{Fx-tails-mmin}
F(x) \sim 
\begin{cases}
x^{n-1}  & x\ll 1\, ,\cr 
\exp\left(-{\rm const.}\times x^\beta\right)      & x\gg 1\,.
\end{cases} 
\end{eqnarray}
The small-$x$ tail is consistent with \eqref{ck-asymp-mmin} and
additionally, it indicates that $A_k\sim k^{n-1}$ when $1\ll k\ll
t$. The suppression of small clusters becomes stronger and stronger as
$n$ grows. In this sense, choice provides a mechanism for controlling
the size distribution. The large-$x$ tail is steeper than
an exponential, and the exponent $\beta$ is determined by 
\begin{equation}
\label{sigma-mmin}
n\sigma^\beta+(1-\sigma)^\beta=1\,,
\end{equation}
along with the selection criterion \eqref{criterion}. Appendix B
provides additional details on the derivation of $\beta$, and Table I
lists several values of $\beta$. Since $\beta$ increases with $n$,
suppression of large clusters becomes stronger with increasing $n$.
 
\section{Symmetric Choice}
\label{sec:symmetric}

In Sects.~\ref{sec:Max}--\ref{sec:many} we implemented choice
asymmetrically: One cluster was selected from the outset, while its
merging partner was chosen from two or more alternatives. Asymmetric
choice can arise, for example, in network growth when a new node
considers a few provisional links, and then implements only one of
these links according to a pre-determined selection criterion. We
recall that the Achlioptas process is symmetric \cite{ASS}, namely two
pairs of nodes are randomly chosen and the link between nodes from one
pair is made. This motivates one to introduce choice using the very
same procedure where clusters from one of the two randomly selected
pairs merge.

\begin{figure}
\centering
\includegraphics[width=5cm]{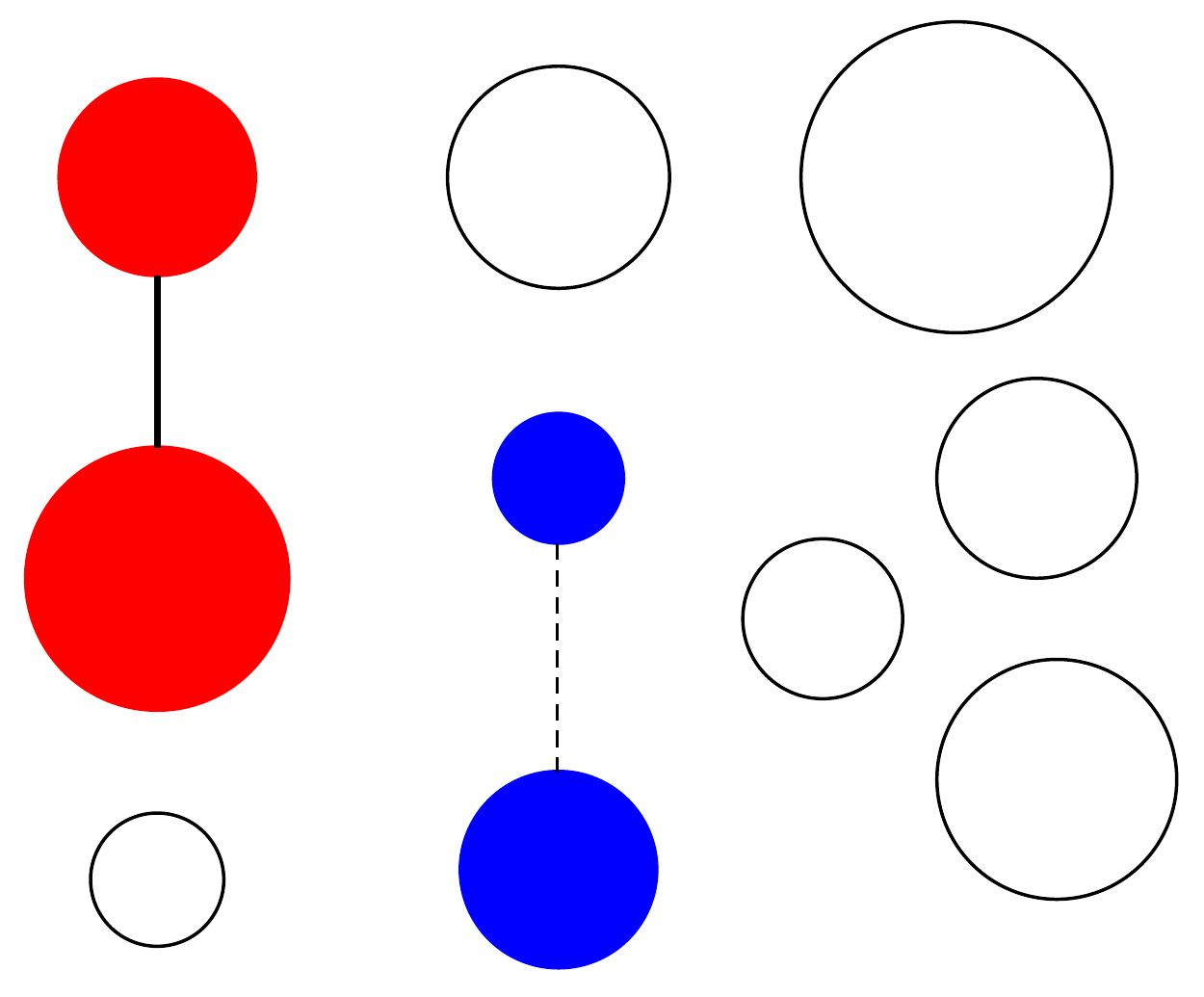}
\caption{Illustration of aggregation with symmetric
  choice. Two pairs (two red disks and two blue disks) are randomly
  drawn. The pair with bigger combined size (red pair) is chosen in the
  maximal choice case. In the minimal choice case, the pair with
  smaller combined size (blue pair) is chosen.}
\label{Fig:pair_choice}
\end{figure}

\subsection{Maximal choice}
\label{max-choice}

In the symmetric version of aggregation with choice, we choose two pairs
of clusters with sizes $i_1,j_1$ and $i_2,j_2$. All four clusters are
chosen randomly.  Without loss of generality, we assume that
$i_1+j_1\geq i_2+j_2$. Under maximal choice, the pair with the
larger total mass undergoes aggregation (see
Fig.~\ref{Fig:pair_choice}):
\begin{equation}
\label{process-smax}
i_1,j_1,i_2,j_2\to i_1+j_1,i_2,j_2\,.
\end{equation}
Hence, the selection criterion is such that the total size of the
resulting aggregate is maximized.  In the Achlioptas process \cite{ASS}, in
contrast, the selection criterion is different, e.g. the {\em product} of
the sizes can be sought to be maximal, so that the choice \eqref{process-smax}
is made if \hbox{$i_1\cdot j_1 \geq i_2\cdot
  j_2$}.

The aggregation process \eqref{process-smax} involves four clusters,
and the corresponding master equation governing the cluster-size
density is quartic 
\begin{eqnarray}
\label{ck-eq-smax}
c^2\,\frac{dc_k}{dt} &=& \sum_{i+j=k}c_ic_j\left(2\sum_{k'<k}c_{i'}c_{j'}
+ \sum_{k'=k}c_{i'}c_{j'}\right) \\
       &-& 2c_k\Bigg(2\sum_{k'<k+j}c_jc_{i'}c_{j'} +\sum_{k'=k+j} c_j c_{i'} c_{j'}\Bigg)\nonumber\,,
\end{eqnarray}
with $k'=i'+j'$. There are two gain terms and two loss terms. The first
gain term accounts for the case where the two pairs have different
total size, and the second gain term, for the complementary case of
equal total size. The two loss terms are similarly ordered.  

\begin{figure}[t]
\includegraphics[width=0.44\textwidth]{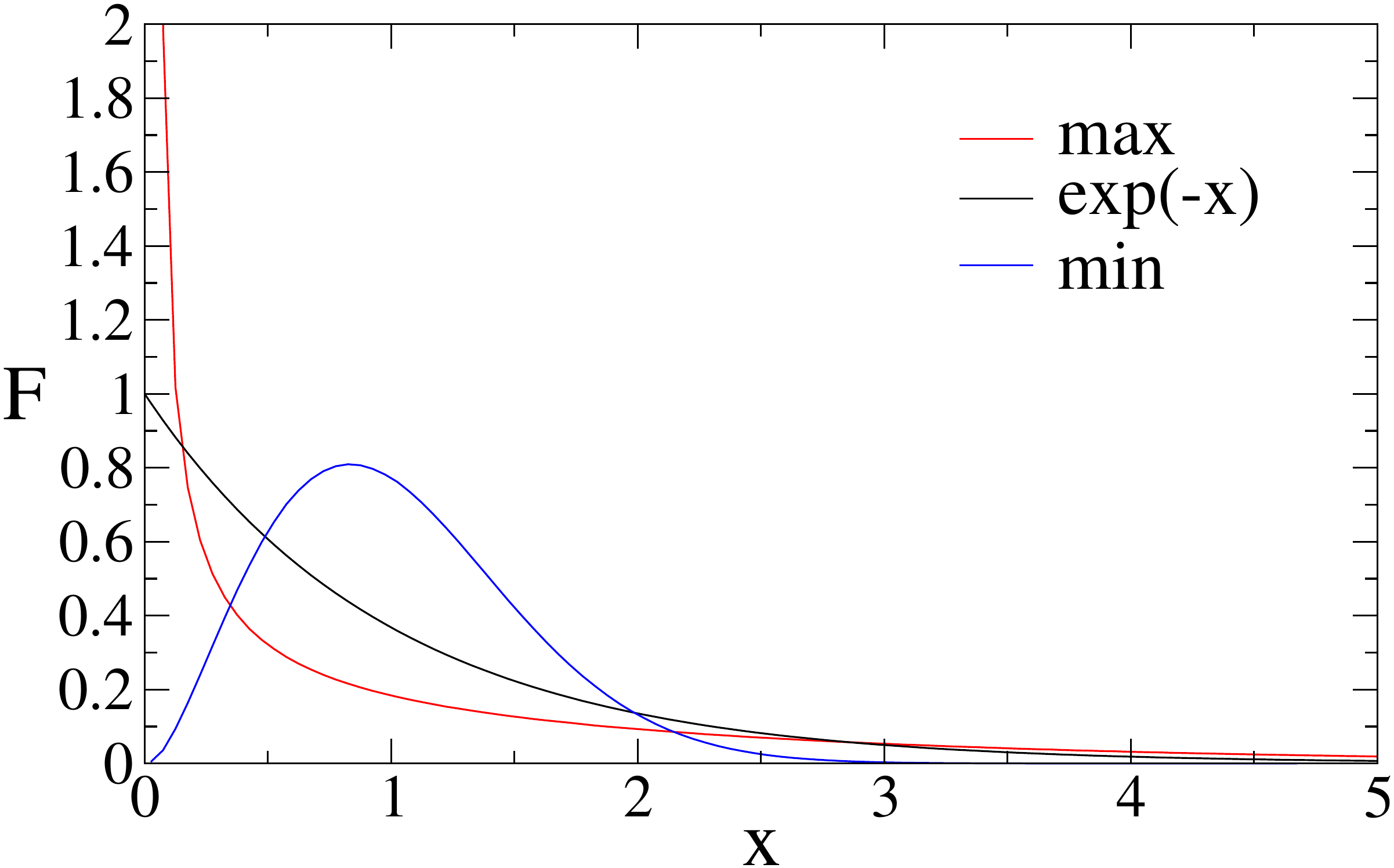}
\caption{The scaling function $F(x)$ versus the scaling variable $x$
  for maximal choice, ordinary aggregation, and minimal choice.}
\label{fig-fx-sym}
\end{figure}

\begin{figure}[t]
\includegraphics[width=0.44\textwidth]{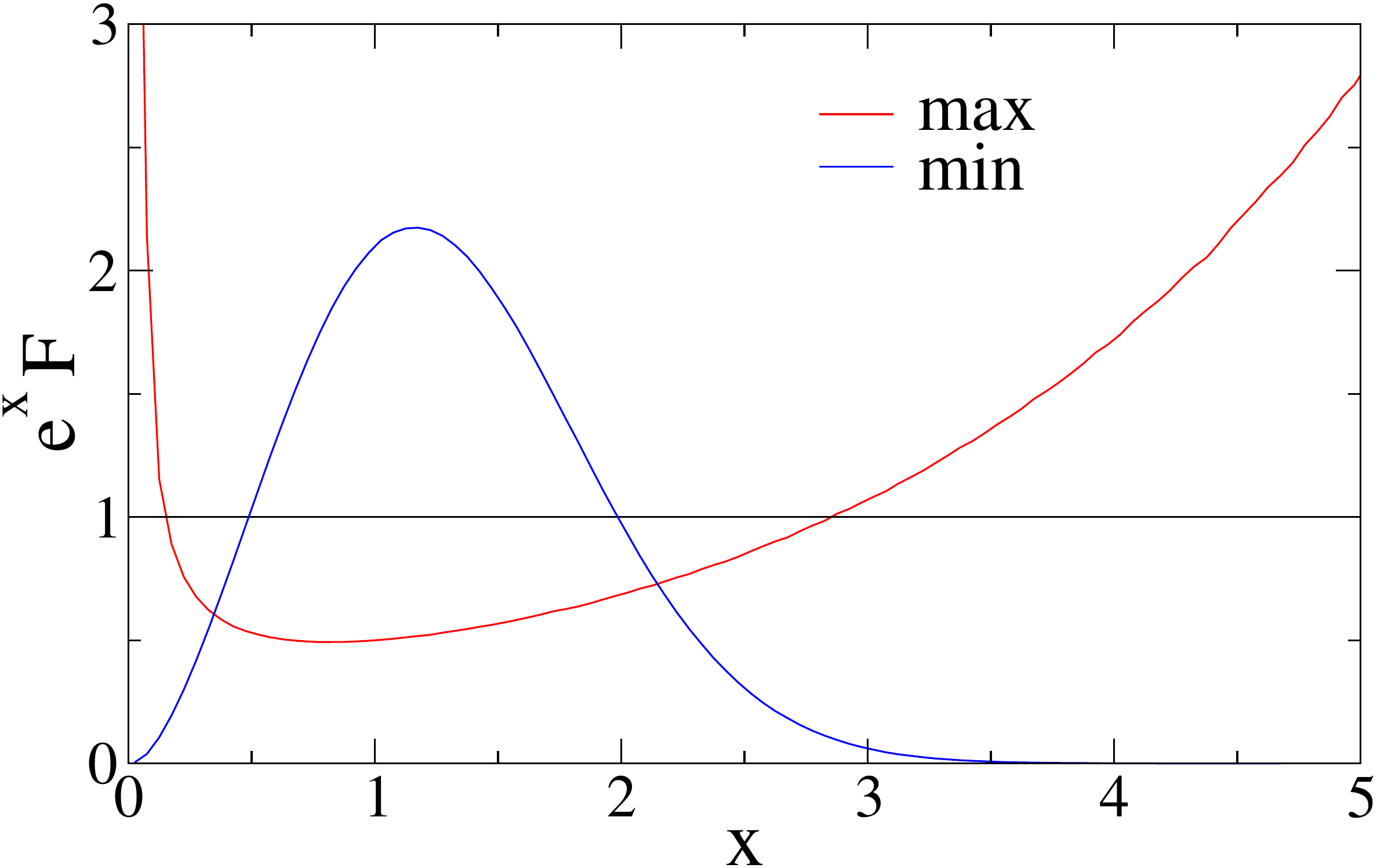}
\caption{The normalized scaling function $e^x\,F(x)$ versus the
  scaling variable $x$.  Shown are the cases of maximal choice,
  ordinary aggregation, and ordinary aggregation.}
\label{fig-fx-sym-compare}
\end{figure}

Our numerical simulations show that the scaling function $F(x)$
maintains the same qualitative features as in the asymmetric case.
Figure \ref{fig-fx-sym} shows that the scaling function $F(x)$
diverges at small-$x$, thereby indicating an overpopulation of small
clusters. Similarly, figure \eqref{fig-fx-sym-compare} which shows
the normalized scaling function $e^{x}F(x)$ demonstrates that there is
also an overpopulation of large clusters. Once again, there are three
size regimes, and at intermediate sizes, the density is suppressed.

By substituting \eqref{F-def} into \eqref{ck-eq-smax}, we see that the
scaling function obeys
\begin{eqnarray}
\label{F-eq-smax}
0=xF'(x)&+&2F(x)+2\phi(x)\int_0^x dy\, F(y)F(x-y)\nonumber\\
        &-&4F(x)\int_0^\infty dy\, F(y)\phi(x+y)\,.
\end{eqnarray}
In deriving this equation we took into account that the second gain
term and the second loss term which correspond to the case where
$i_1+j_1=i_2+j_2$ are asymptotically negligible. The function
$\phi(z)$ appearing in \eqref{F-eq-smax} is the shorthand notation for
the following integral
\begin{eqnarray}
\label{phi-def}
\phi(z)=\iint_{x'+y'<z}dx'dy'F(x')F(y')\,.
\end{eqnarray}
The scaling function is subject to the normalization
\eqref{F-int}. 

At small sizes, the convolution term in \eqref{F-eq-smax} is
negligible and it simplifies to $xF'=(\gamma-2)F$ with 
\begin{eqnarray}
\label{gamma-smax}
\gamma=4\int_0^\infty dy\, F(y)\phi(y)\,.
\end{eqnarray}
The scaling function is therefore algebraic, $F(x)\sim
x^{\gamma-2}$, when $x\ll 1$. This algebraic behavior implies the
algebraic decay 
\begin{eqnarray}
\label{ck-asymp-smax}
c_k\sim t^{-\gamma}
\end{eqnarray}
for finite $k\ll t$. Indeed, it is possible to derive
\eqref{ck-asymp-smax} with \eqref{gamma-smax} directly from the master
equation \eqref{ck-eq-smax} together with the scaling form
\eqref{F-def}. The behavior \eqref{ck-asymp-smax} also holds for
monomers, and there is no longer an anomaly associated with minimal
clusters.

The exponent $\gamma$, which according to
\eqref{gamma-smax} requires full knowledge of 
$F(x)$, appears to be nontrivial. Our numerical simulations yield
$\gamma=1.25\pm 0.01$ (Fig.~\ref{fig-c1-sym}). If we ignore the
logarithmic correction in \eqref{Fx-smallx-max}, then the
corresponding value for the asymmetric case is 
$\gamma=1$. We have not determined $\gamma$ analytically, but 
in Appendix D we derive the bounds 
\begin{equation}
\label{gamma-bounds-smax}
1\leq \gamma<\frac{4}{3}\,.
\end{equation}
According to these bounds, the scaling function diverges in the limit
$x\to 0$ (see Fig.~\ref{fig-fx-sym}). 

At large sizes, the convolution term dominates and $\phi\to 1$, so
that Eq.~\eqref{F-eq-smax} simplifies to
\eqref{Fx-max-large}. Consequently, $F(x)$ decays exponentially
according to \eqref{Fx-largex-max}. Numerically, we find the decay
constant $\alpha=0.53\pm 0.01$, which is slightly smaller than the
value $\alpha=0.57\pm 0.01$ for the asymmetric case. The extremal
behaviors of the scaling function are therefore
\begin{equation}
\label{F-tails-smax}
F(x)\sim 
\begin{cases}
x^{\gamma-2}              &x\to 0,\\
e^{-\alpha x}                &x\to \infty.\\
\end{cases}
\end{equation}

Thus, many of the features obtained for aggregation with asymmetric
choice extend to aggregation with symmetric choice. The density of
very small and very large clusters are enhanced at the expense of
moderate-size clusters. The normalized size density again diverges at
small sizes, and interestingly, this divergence is characterized by a
nontrivial exponent. There is one difference between the two cases,
however. The scaling function captures the asymptotic behavior at all
scales and there is no anomaly associated with small clusters.

\subsection{Minimal choice}
\label{min-choice}

We now consider the complementary case where the pair with the minimal
total mass undergoes aggregation. Aggregation proceeds according
to \eqref{process-smax} except that now $i_1+j_1\leq i_2+j_2$. Repeating the 
above analysis one finds that the
scaling function satisfies 
\begin{eqnarray}
\label{F-eq-smin}
0=xF'(x)&+&2F(x)+2\psi(x)\int_0^x dy F(y)F(x-y)\nonumber\\
        &-&4F(x)\int_0^\infty dy F(y)\psi(x+y)\,.
\end{eqnarray}
This equation differs from \eqref{F-eq-smax} in that 
$\phi(y)$ is replaced by the complementary integral 
\begin{eqnarray}
\label{psi-def}
\psi(y)=\iint_{x'+y'>y}dx'dy'F(x')F(y')\,,
\end{eqnarray}
so that $\psi(y)+\phi(y)=1$ for all $y$.  Asymptotic analysis of
equation \eqref{F-eq-smin} yields
\begin{equation}
\label{F-tails-smin}
F(x)\sim 
\begin{cases}
x^{\gamma-2}                       &x\to 0,\\
e^{-{\rm const.}\times x^2}   &x\to \infty.\\
\end{cases}
\end{equation}

\begin{figure}
\includegraphics[width=0.44\textwidth]{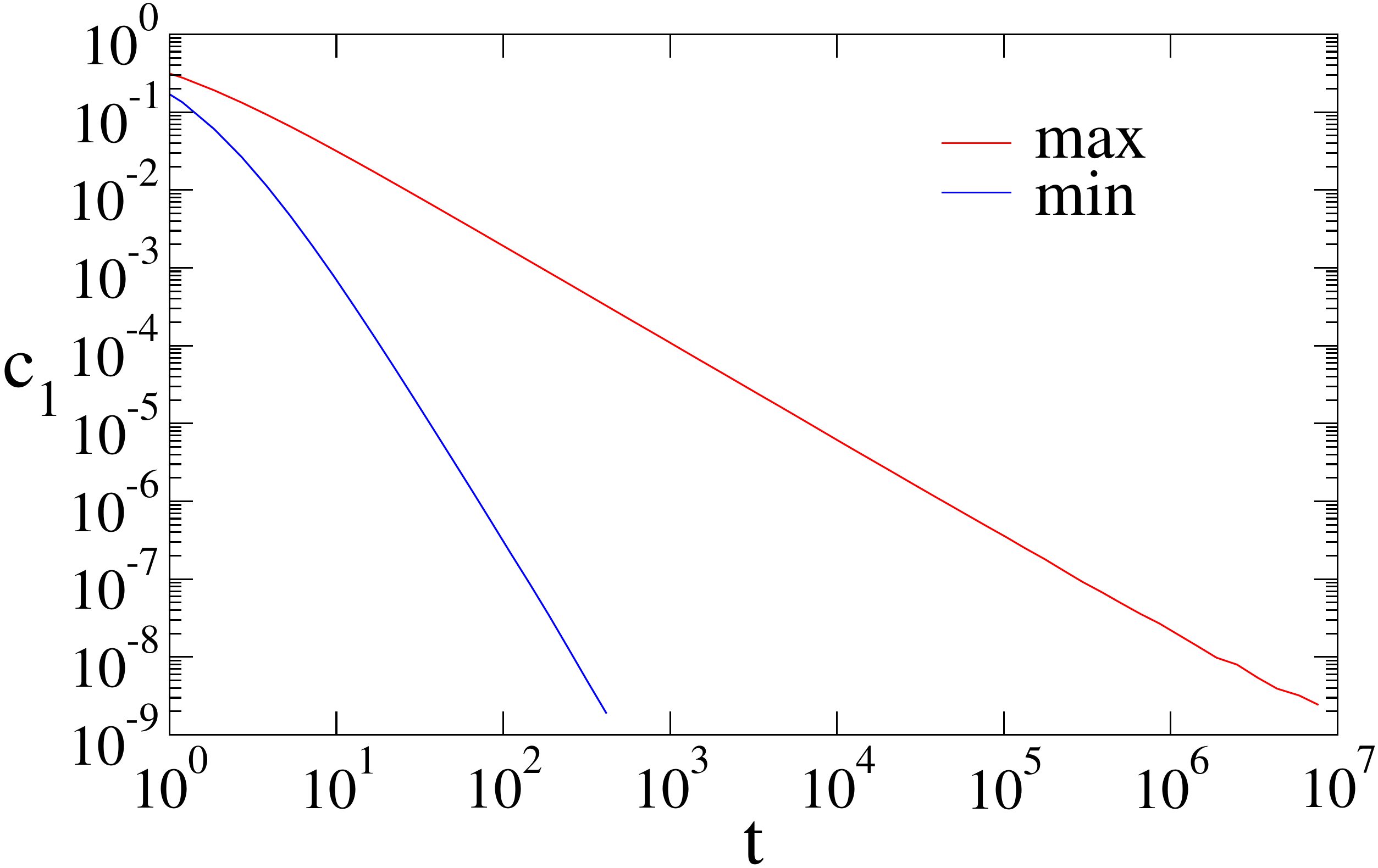}
\caption{The monomer density $c_1(t)$ versus time $t$ for symmetric
  aggregation with maximal and minimal choice.}
\label{fig-c1-sym}
\end{figure}

The small-$x$ behavior is characterized by the nontrivial exponent
$\gamma$ which is given by the analog of \eqref{gamma-smax}
\begin{equation}
\gamma=4\int_0^\infty dy\, F(y)\psi(y)\,.
\end{equation}
Numerically, we find the value $\gamma=3.5\pm 0.1$
(Fig.~\ref{fig-c1-sym}), which is somewhat larger than the value
$\gamma=3$ above. Hence, the suppression of small clusters becomes
stronger under the symmetric aggregation process \eqref{process-smax}.
In Appendix D, we obtain the bounds
\begin{equation}
\label{gamma-bounds-smin}
\frac{8}{3}\leq \gamma \leq 4\,.
\end{equation}
The small-$x$ tail \eqref{F-tails-smin} implies that the density of
small clusters decay algebraically with time according to
\eqref{ck-asymp-smax}.

To estimate the large-size tail, we first note that for a
sharply-decaying $F(x)$, the integrand in \eqref{psi-def} is maximal
at $x'=y'=x/2$, and as a result $\psi(x)\asymp F^2(x/2)$. Following
this reasoning, we estimate that the convolution term in
\eqref{F-eq-smin} behaves as $F^2(x/2)$. For large $x$, the derivative
term and the convolution term dominate, and balancing these two terms
gives
\begin{equation}
\label{F-largex-eq-smin}
-xF'(x)\asymp F^4(x/2)\,.
\end{equation}
We now substitute the super-exponential form \eqref{Fx-largex-min} and
obtain $1=4/2^\beta$ from which we deduce $\beta=2$ leading to the
Gaussian tail in \eqref{F-tails-smin}. Compared with the value
$\beta=1.26749$ in the asymmetric case, we deduce that the tail is now 
sharper.

Figures \ref{fig-fx-sym} and \ref{fig-fx-sym-compare} compare the
scaling function $F(x)$ with the exponential decay $e^{-x}$ which
corresponds to ordinary aggregation. As in the case of symmetric
aggregation, the populations of very small and very large clusters are
suppressed, while the population of intermediate clusters is enhanced.
we conclude that the qualitative behavior of the size density for
aggregation with symmetric and asymmetric choice are similar.

\section{Conclusions}

In summary, we generalized the most basic aggregation process to
include choice. In our implementation, several clusters are drawn at
random, and two clusters merge while the rest are not affected. The
merging clusters are chosen in a way that maximizes or minimizes the
aggregate size. We considered several versions and found a number of
common features. In all cases, the size density adheres to standard
scaling, in contrast with some aggregation processes in which scaling
is violated (see e.g. \cite{ASP82,HE84,LR87,MKR03}.)

In general, introduction of choice changes the shape of the
cluster-size distribution. When the merger maximizes the size of the
final aggregate, the small-size tail of the distribution is enhanced
because small clusters are less likely to undergo
aggregation. Surprisingly, the large-size tail of the distribution is
also enhanced. The opposite effect emerges when the merging clusters
minimize the aggregate size. These qualitative features are general,
and hold regardless of the number of clusters involved in the
aggregation process.

We found a number of interesting features for aggregation with
choice. In the asymmetric version with maximal choice, the scaling
function does not capture the entire size density.  In particular,
when $n$ clusters are involved in the aggregation process, there are
$n$ distinct scaling laws that characterize the density of monomers,
dimers, up to $n$-mers.  The population of these small clusters is
anomalously large compared with that of typical clusters.  In the
asymmetric version with minimal choice, the large-$x$ tail is
super-exponential $F(x)\sim \exp[-{\rm const.}\times x^\beta]$, and it
is governed by a nontrivial exponent $\beta>1$.  This exponent is
selected from a spectrum of possible values according to a principle
that is reminiscent of velocity selection in nonlinear traveling
waves.

One model with symmetric choice which would be interesting to explore
is the following: Pick up randomly three clusters and merge two of
them, e.g. the smallest or the largest. We analyzed our models only in
the mean-field case, and another extension is to aggregation in finite
spatial dimensions. For instance, clusters may occupy a single lattice
site, and hop to adjacent sites with the same mass-independent rate,
and when three clusters occupy the same site, two of them, say the
smallest, merge. The third (largest) cluster thus plays a role of a
catalyst.  This toy model is inspired by living matter, as most
biological processes involve a catalyst.

\medskip\noindent 

We acknowledge support from US-DOE grant DE-AC52-06NA25396 (EB).

\appendix
\section{The fraction of dimers $C_2$}
\label{app:max}

For maximal choice, the fraction of dimers obeys
\begin{equation}
\label{C2-eq-max}
\frac{dC_2}{d\tau}= C_1^3 + C_1^2 - (C_1+C_2)^2\,.
\end{equation}
Using $C_1=1/(1+\tau)$, we obtain the Riccati equation
\begin{equation}
\label{C2-eq-R-max}
\frac{dC_2}{d\tau}=-C_2^2-\frac{2}{1+\tau}C_2+\frac{1}{(1+\tau)^3}\,.
\end{equation}
To find the solution we first linearize the first-order nonlinear
differential equation \eqref{C2-eq-R-max} by making the transformation
$C_2=-[2u^3f(u)]/f'(u)$ with $u=(1+\tau)^{-1/2}$. The quantity $f(u)$ 
obeys the Bessel equation, and thereby, we arrive at the dimer fraction 
\begin{equation}
\label{C2-max}
C_2=u^3\,\frac{I_0(2)K_0(2\,u)-K_0(2)I_0(2\,u)}{I_0(2)K_1(2\,u)+K_0(2)I_1(2\,u)}.
\end{equation}

For minimal choice, we have 
\begin{equation}
\label{C2-eq-min}
\frac{dC_2}{d\tau} = 2C_1^2 +C_2^2-C_1^3+2C_1C_2-2C_2\ .
\end{equation}
We now write 
\begin{equation}
\label{C2-transform}
C_2=e^{-2\tau} U_2(T), \quad T = e^{-2\tau}.
\end{equation}
Recalling that $C_1=2/(1+e^{2\tau})=2T/(1+T)$ and using \eqref{C2-transform} we recast \eqref{C2-eq-min} into
\begin{equation}
\label{C2-eq-R-min}
\frac{dU_2}{dT}=-\frac{1}{2}U_2^2- \frac{2}{1+T} \,U_2 - \frac{4}{(1+T)^3}
\end{equation}
This Riccati equation should be solved subject to the initial
condition $C_2(T=1)=0$. We use the same procedure as before: We
linearize \eqref{C2-eq-R-min} by making the transformation
$U_2=[v^3f(v)]/[8f'(v)]$ with $v=\sqrt{8/(1+T)}$. Again, the function
$f(v)$ obeys the Bessel equation, and 
\begin{equation}
\label{U2-min}
U_2 = \frac{v^3}{8}\,\frac{J_0(2)Y_0(v)-Y_0(2)J_0(v)}{Y_0(2)J_1(v)-J_0(2)Y_1(v)}\,.
\end{equation}
By combining \eqref{C2-transform} and \eqref{U2-min}, we arrive at the
announced result \eqref{c2-min} for the dimer density.

\section{The exponent $\beta$}
\label{app:beta}

To determine the large mass decay in the minimal choice model, we must
solve \eqref{sigma} and \eqref{criterion}. We explain the procedure in
the general case of $n$ alternatives.  Let us fix $n>1$ and examine
$\sigma$ as a function of $\beta$.  The derivative $d\sigma/d\beta$ reaches maximum at a single point.
Indeed, $\sigma(\beta)$ is a
monotonically increasing function which sharply vanishes when
$\beta\to 1$ and algebraically approaches unity when $\beta\to
\infty$, that is,
\begin{equation}
\sigma\to
\begin{cases}
n^{-1/(\beta-1)}      & \beta\to 1,\\
1-\beta^{-1}\ln n   & \beta\to \infty. 
\end{cases}
\end{equation}
Thus we seek a solution to Eqs.~\eqref{criterion} and \eqref{sigma-mmin}. The explicit form of the former equation is rather
cumbersome,
\begin{eqnarray}
\label{crit}
\frac{2}{\beta} &=&\frac{(1 - \sigma)^\beta [\ln(1 - \sigma)]^2 + n \sigma^\beta [\ln \sigma]^2}{S}  \nonumber\\
&+& 2\,\frac{n \sigma^{\beta - 1} \ln \sigma 
    - (1 - \sigma)^{\beta - 1} \ln(1 - \sigma)}{(1 - \sigma)^{\beta - 1} -
     n \sigma^{\beta - 1}}  \nonumber \\
&+& S\,\frac{\beta-1}{\beta}\, \frac{(1 - \sigma)^{\beta - 2} + 
    n \sigma^{\beta - 2}}{[(1 - \sigma)^{\beta - 1} - n \sigma^{\beta - 1}]^2}
\end{eqnarray}
where $S = (1 - \sigma)^\beta \ln(1 - \sigma) + n \sigma^\beta
\ln\sigma$. The two transcendental equations, \eqref{sigma-mmin} and
\eqref{crit}, can be solved using e.g. {\em Mathematica}.

\section{The Amplitude $A_1$}
\label{app:AB}

Here, we list explicit expressions for $\ln A_1$ for $n\leq 6$ 
\begin{equation*}
\ln A_1=\begin{cases}
0 & n=1 \\
\ln 2 & n=2, \\
\frac{\pi \sqrt{3}}{6} + \frac{1}{2}\,\ln 3 & n=3,\\
\frac{\pi}{2} + \ln 2 & n=4, \\
\frac{\pi}{2}\sqrt{1+\frac{2}{\sqrt{5}}}  + \frac{\sqrt{5}}{2}\,\arctan\left[\frac{1}{\sqrt{5}}\right]  + \frac{1}{4}\,\ln 5 & n=5,\\
\frac{\pi \sqrt{3}}{2} + \frac{1}{2}\,\ln 3 + \ln 2 & n=6. \\
\end{cases}
\end{equation*}
In particular, when $n=2$ we recover $A_2=2$, consistent with the
exact solution \eqref{c1-min}.

\section{The exponent $\gamma$}
\label{app:bounds}

First, we derive the bounds \eqref{gamma-bounds-smin}.  The quantity
$\psi(y)$ defined in \eqref{psi-def} is monotonically decreasing and
since $\psi(0)=1$ we have $\psi(y)\leq 1$. The upper bound readily
follows: 
\begin{equation}
\label{gamma-upper-smin}
\gamma=4\int_0^\infty dy\,F(y)\psi(y)\leq 4\int_0^\infty dy\,F(y)=4.
\end{equation}

To derive the lower bound, we narrow the integration range in the
integral in \eqref{psi-def} from $x'+y'>y$ to the union of the
vertical strip $0<x'<y, ~y'>y$, the horizontal strip $x'>y, ~0<y'<y$,
and the quadrant $x'>y, ~y'>y$. The contribution to $\psi(y)$ from the
vertical strip is
\begin{equation}
\label{strip}
\int_0^y dx'\,F(x')\int_y^\infty dy'\,F(y')=[1-\Psi(y)]\Psi(y),
\end{equation}
where $\Psi(y)=\int_y^\infty dz\,F(z)$.  The contribution from the
horizontal strip is also given by Eq.~\eqref{strip}. The contribution
to $\psi(y)$ from the quadrant $x'>y, ~y'>y$ is $\Psi^2(y)$. Summing
these contributions we obtain
\begin{equation}
\psi(y)\geq \Psi^2(y)+2[1-\Psi(y)]\Psi(y)=2\Psi(y)-\Psi^2(y).
\end{equation}
The lower bound is obtained as follows 
\begin{eqnarray}
\label{gamma-lower-smin}
\gamma  &\geq& 4\int_0^\infty dy\left(-\frac{d\Psi}{dy}\right)[2\Psi(y)-\Psi^2(y)]\nonumber\\
            &=& 4\int_0^1 d\Psi\,(2\Psi-\Psi^2) = \frac{8}{3}.
\end{eqnarray}

To establish the upper bound in Eq.~\eqref{gamma-bounds-smax} we
extend the integration range in $\phi(y)$, Eq.~\eqref{phi-def}, from
the triangle $x'+y'<y$ to the square $0<x',y'<y$. This gives
$\phi(y)\leq [1-\Psi(y)]^2$, and therefore,
\begin{eqnarray}
\gamma \leq 4\int_0^\infty dy\left(-\frac{d\Psi}{dy}\right)[1-\Psi(y)]^2 = \frac{4}{3}.
\end{eqnarray}
The lower bound in \eqref{gamma-bounds-smax} follows from $c_1(t)\leq c(t)$.

\end{document}